\documentclass[runningheads]{llncs}
\usepackage{calc}
\usepackage{tikz}
\usetikzlibrary{decorations.pathreplacing}
\usetikzlibrary{decorations.markings}
\tikzstyle{vertex}=[circle, draw, inner sep=0pt, minimum size=3pt]

\usepackage{graphicx}
\usepackage{amsmath}
\usepackage{amsfonts}
\usepackage{xcolor}
\usepackage{amssymb}
\usepackage{graphicx}
\usepackage{tikz}
\usetikzlibrary{circuits.logic.US}
\usetikzlibrary{positioning}
\usetikzlibrary{matrix}
\usepackage{graphicx}
\usepackage{amsmath}
\newcommand{\boundellipse}[3]
{(#1) ellipse (#2 and #3)
}
\definecolor{magenta}{rgb}{0.8, 0.0, 0.8}
\definecolor{cyan}{rgb}{0.0, 1.0, 1.0}
\definecolor{green1}{rgb}{0.1, 0.6, 0.01}
\definecolor{green}{rgb}{0.11, 0.35, 0.02}
\definecolor{brown}{rgb}{0.65, 0.16, 0.16}
\definecolor{cadetgrey}{rgb}{0.57, 0.64, 0.69}

\newtheorem{red}{\bf Reduction UEDS}

\title{Parameterized Complexity of Upper Edge Domination}
\titlerunning{The Upper Edge Domination Problem}
\author{Ajinkya Gaikwad
 \and Soumen Maity
}
\authorrunning{A.\,Gaikwad and S.\,Maity}
\institute{Indian Institute of Science Education and Research, Pune, India 
\email{\texttt{ajinkya.gaikwad@students.iiserpune.ac.in;}}
\email{\texttt{soumen@iiserpune.ac.in}}
}

\begin{document}

\maketitle          

\begin{abstract}
   In this paper we study a maximization version of 
the classical {\sc Edge Dominating Set (EDS)} problem, namely, the {\sc Upper EDS} problem, in the realm of Parameterized Complexity. In this problem, given an undirected graph $G$, a positive integer $k$, the question is to check whether $G$ has a minimal edge
dominating set of size at least $k$.  We obtain the following results for {\sc Upper EDS}. 
We prove that {\sc Upper EDS} admits a kernel  with at most $4k^2-2$ vertices. 
We also design  a fixed-parameter tractable (FPT)  algorithm for  
    {\sc Upper EDS}  running in time $2^{\mathcal{O}(k)} \cdot n^{\mathcal{O}(1)}$.
\end{abstract}

\section{Introduction}
The dominating set problem and its variants have been extensively studied in the literature. 
Typically, researchers have considered this concept in terms of the minimisation problem
{\sc Minimum Dominating Set}, namely  {\sc Min DS}: find a smallest set of vertices that dominate all vertices of the graph \cite{domination}. However, researchers have also considered the max-min variant, usually called {\sc Upper Dominating Set}, which we abbreviate to {\sc Upper DS}: A minimal dominating set is a dominating set in a graph that is not a proper subset of any other dominating set. Every minimum dominating set is a minimal dominating set, but the converse does not necessarily hold. Our goal here is to find an inclusion-wise minimal dominating set of largest size  \cite{COC,CHESTON1990195,JACOBSON199059,Fellows1994ThePN,COCOA-Domination,IWOCA-Domination,Bazgan-Domination,Bazgan-Domination2,BAZGAN-Domination3,BAZGAN-Domination4}. 
Both {\sc Min DS} and 
{\sc Upper DS} are NP-hard for general graphs; see [\cite{Garey}, problem GT2] and \cite{CHESTON1990195}, respectively. In 2021,  Monnot, Fernau and Manlove \cite{MONNOT202146} studied the edge variant of the (vertex) dominating set problem. 

\par A set of edges $M$ of $G=(V,E)$ is called an \emph{edge dominating set} if every edge of $E\setminus M$ is adjacent to some edge of $M$. 
Similarly, researchers have considered this concept in terms of the minimisation problem
{\sc Minimum Edge Dominating Set}, namely  {\sc Min EDS}: find a smallest set of edges that dominate all edges of the graph \cite{SIAM-ED}.
An edge dominating set $M$ of
$G$ is said to be a \emph{minimal edge dominating set} if no proper subset of $M$ is also
an edge dominating set of $G$.
The problem We consider in this paper is as follows:
\vspace{3mm}
    \\
 \noindent   \fbox
    {\begin{minipage}{37.7em}\label{FFVS }
       {\sc Upper EDS}\\
        \noindent{\bf Input:} A graph $G=(V,E)$ and an integer $k$.\\
     \noindent{\bf Question:}  Does $G$ have a minimal edge dominating set $M\subseteq E$ of size at least $k$?
    \end{minipage} }
    \vspace{3mm}
    \\
Whilst {\sc Min EDS} has received considerable attention in the literature, the same is not true for {\sc Upper EDS}.
{\sc Min EDS} is NP-hard in planar or bipartite graphs of maximum degree 3 \cite{Yan-ED} and in planar cubic graphs \cite{SIAM-ED}, whilst solvable in polynomial time in several graph classes (see \cite{JCO} for a brief survey). 
On the other hand {\sc Upper EDS} has been largely neglected: {\sc Upper EDS} is NP-hard in bipartite graphs \cite{mcrae1994generalizing}.
Monnot et al. \cite{MONNOT202146} showed that this problem is not approximable
within a ratio of $n^{\epsilon -\frac{1}{2}}$ , for any $\epsilon \in (0,1)$, assuming $P\neq NP$, where $n=|V|$. In this paper we enhance our understanding of the problem from the viewpoint of parameterized complexity. 
 We refer to \cite{marekcygan,Downey} for further details on parameterized complexity. \\
\noindent Our results are as follows:
\begin{itemize}
 \item  {\sc Upper Edge Dominating Set}  parameterized by the solution size $k$ admits a kernel of size $4k^{2}-2$.
 \item We prove that, given an $n$-vertex graph $G$ and its nice tree decomposition $T$ of width at most $\omega$, the 
  size of a maximum {\sc Upper Edge Dominating Set} of $G$ can be computed in time $45^{\omega}\cdot n^{\mathcal{O}(1)}$. This gives a fixed-parameter tractable (FPT)  algorithm for  
    {\sc Upper EDS}  running in time $2^{\mathcal{O}(k)} \cdot n^{\mathcal{O}(1)}$
 \end{itemize}

\section{Preliminaries} Throughout this paper, we consider simple undirected graphs. 
A graph  $G=(V,E)$ can be specified by the set $V$ of vertices and the set $E$ of edges.
 The {\it (open) neighbourhood} $N_G(v)$ of a vertex 
$v\in V$ is the set $\{u~|~(u,v)\in E\}$. The {\it closed neighbourhood} $N_G[v]$ of a vertex $v\in V$ is the set
$\{v\} \cup N_G(v)$.  The subgraph induced by 
$D\subseteq V$ is denoted by $G[D]$.
Every edge has two endpoints and these two endpoints are called \emph{adjacent}; if $v$ is an endpoint of $e$, we also say that $e$ and $v$ are  \emph{incident} and two edges $e$ and $e'$ are adjacent if they share a common endpoint.
The {\it (closed) neighbourhood} $N[e]$ of an edge
$e\in E(G)$ is the set $\{e'~|~e \mbox{ and } e' \mbox{ are adjacent}\}$.

An edge set $M\subseteq E$ is an \emph{edge dominating set} if every edge $e \in  E \setminus M$ is adjacent to some edge of $M$. 
Let $M \subseteq E$ be an edge dominating set. Define an edge $e \in E$ to be \emph{private} if $e$ is dominated by exactly one edge of $M$.
The following lemma demonstrates a connection between minimal edge dominating sets and private edges

\begin{lemma} \cite{MONNOT202146} \label{private-edge-lemma}\rm 
Let $G = (V , E)$ be a graph and let $M \subseteq E$ be an edge dominating set. 
Then $M \subseteq E$ is a minimal edge dominating set if and only if every edge $e \in M$ has a private edge in $N[e]$.
\end{lemma}

The graph
parameter that we explicitly use in this paper is  treewidth.
We  review the concept of a tree decomposition, introduced by Robertson and Seymour in \cite{Neil}.
Treewidth is a  measure of how “tree-like” the graph is.
\begin{definition}\rm \cite{Downey} A {\it tree decomposition} of a graph $G=(V,E)$  is a tree $T$ together with a 
collection of subsets $X_t$ (called bags) of $V$ labeled by the vertices $t$ of $T$ such that 
$\bigcup_{t\in T}X_t=V $ and (1) and (2) below hold:
\begin{enumerate}
			\item For every edge $uv \in E(G)$, there  is some $t$ such that $\{u,v\}\subseteq X_t$.
			\item  (Interpolation Property) If $t$ is a vertex on the unique path in $T$ from $t_1$ to $t_2$, then 
			$X_{t_1}\cap X_{t_2}\subseteq X_t$.
		\end{enumerate}
	\end{definition}
	
\begin{definition}\rm \cite{Downey} The {\it width} of a tree decomposition is
the maximum value of $|X_t|-1 $ taken over all the vertices $t$ of the tree $T$ of the decomposition.
The treewidth $tw(G)$ of a graph $G$  is the  minimum width among all possible tree decompositions of $G$.
\end{definition} 
\noindent A special type of tree decomposition, known as a {\it nice tree decomposition}, was 
introduced by Kloks \cite{Kloks1994TreewidthCA}.  The nodes in such a decomposition can be partitioned into four types:

\begin{definition}\rm \cite{Kloks1994TreewidthCA} A tree decomposition is said to be a {\it nice tree decomposition} if the following conditions are satisfied:
	\begin{enumerate}
		\item All bags that correspond to leaves are empty. One of the leaves is considered as root node $r$. Thus $X_r=\emptyset$ and $X_{l}=\emptyset$ for each leaf $l$. 
		\item There are three types of non-leaf nodes:
		\begin{itemize}
		 \item {\bf Introduce node:} a node $t$ with exactly one child $t^{\prime}$
		such that $X_{t}=X_{t^{\prime}} \cup \{v\}$ for some $v \notin X_{t^{\prime}}$;
		we say that $v$ is {\it introduced} at $t$.
		\item {\bf Forget node:} a node $t$ with exactly one child $t^{\prime}$
		such that $X_{t}=X_{t^{\prime}} \setminus \{w\}$ for some $w \in X_{t^{\prime}}$;
		we say that $w$ is {\it forgotten}  at $t$.
		\item {\bf Join node:} a node with two children $t_1$ and $t_2$ such that $X_t=X_{t_1}=X_{t_2}$.
		\end{itemize}
	\end{enumerate}
\end{definition}
\noindent Note that, by the interpolation property of tree decomposition, a vertex $v\in V(G)$ may be introduced several times, but each vertex is forgotten  only once. To control the introduction of edges, sometimes one more type of node is considered in a nice tree decomposition called introduce edge node. An {\it introduce edge node} is a node $t$, labeled with edge $uv \in E(G)$, such that $u,v\in X_{t}$ and $X_{t}=X_{t^{\prime}}$, where $t^{\prime}$ is the only child of $t$. We say that node $t$ introduces edge $uv$. We additionally require that every edge of $E(G)$ is introduced exactly once in the whole decomposition.
It is known that if a graph	$G$ admits a tree decomposition of width at most {\tt tw}, then it also admits	a nice tree decomposition of width at most {\tt tw}, that has at most 
	$O(n\cdot {\tt tw})$ nodes \cite{marekcygan}.\\

\section{Kernelization  algorithm  for {\sc Upper EDS} parameterized by 
 solution size}

In this section we give a kernelization algorithm for {\sc Upper EDS} which matches the lower bound. We start with some 
simple reduction rules that clean up the graph.  The first reduction rule is based on the following trivial observation: If the graph $G$ has an isolated vertex, the removal of this vertex does not change the solution, and this operation can be implemented in polynomial time. 
Thus, the following rule is safe. 

\begin{red}\label{Red1}\rm 
If  $G$ contains an isolated vertex $v$, remove $v$ from $G$. 
The resulting instance is $(G-v,k)$. 
\end{red}

\noindent The second rule is also based on a simple observation. If $G$ contains an isolated edge, it must be included in the solution.

\begin{red}\label{Red2}\rm 
If there is an isolated edge $(u,v)$ in $G$, delete it and decrease $k$ by 1. 
The new instance is  $(G-\{u,v\},k-1)$. 
\end{red}

\noindent In our kernelization algorithm, it is convenient to work with coloured graphs.
We colour the vertices of $G$ with four colours: blue, purple, red and green. 
The meaning of the colours is the following. The vertices of degree 1
are coloured blue;
every vertex that is adjacent to a blue  vertex is coloured purple; we colour a vertex red if all of its neighbours are coloured purple; and  rest of the vertices are coloured green.  We denote the set of blue, purple, red and green vertices by $V_{B},V_{P},V_{R}$ and $V_{G}$ respectively. We make a simple note that every green vertex has at least one green neighbour. Also it is easy to verify that this is a valid partition, that is, every vertex is coloured with exactly one colour. Based on the colouring, we give some simple reduction rules.  

\begin{figure}
    \centering
    \begin{tikzpicture}[scale=0.8]
\node[circle,fill=green, draw, inner sep=0 pt, minimum size=0.2cm](c1) at (0,0) [label=below:$g_2$]{};
\node[circle,fill=green, draw, inner sep=0 pt, minimum size=0.2cm](c2) at (1,0) [label=below:$g_3$]{};
\node[circle,fill=green, draw, inner sep=0 pt, minimum size=0.2cm](c3) at (2,0) [label=below:$g_4$]{};
\node[circle,fill=red, draw, inner sep=0 pt, minimum size=0.2cm](c4) at (3,0) [label=below:$r_1$]{};
\node[circle,fill=red, draw, inner sep=0 pt, minimum size=0.2cm](c5) at (4,0) [label=below:$r_2$]{};
\node[circle,fill=green, draw, inner sep=0 pt, minimum size=0.2cm](b1) at (0.5,1) [label=left:$g_1$]{};
\node[circle,fill=magenta, draw, inner sep=0 pt, minimum size=0.2cm](b2) at (2,1) [label=left:$p_1$]{};
\node[circle,fill=magenta, draw, inner sep=0 pt, minimum size=0.2cm](b3) at (3,1) [label=left:$p_2$]{};
\node[circle,fill=magenta, draw, inner sep=0 pt, minimum size=0.2cm](b4) at (4,1) [label=left:$p_3$]{};

\node[circle,fill=cyan, draw, inner sep=0 pt, minimum size=0.2cm](a1) at (2,2) [label=left:$b_1$]{};
\node[circle,fill=cyan, draw, inner sep=0 pt, minimum size=0.2cm](a2) at (3,2) [label=left:$b_2$]{};
\node[circle,fill=cyan, draw, inner sep=0 pt, minimum size=0.2cm](a3) at (4,2) [label=left:$b_3$]{};

\draw(a1)--(b2);
\draw(a2)--(b3);
\draw(a3)--(b4);
\draw(c3)--(b2);
\draw(c4)--(b3);
\draw(c5)--(b4);
\draw(c2)--(b2);
\draw(c2)--(c1);
\draw(c2)--(b1);
\draw(b1)--(c1);
\draw(c2)--(c3);
\draw(c4)--(b2);
\draw(c5)--(b3);
\draw(b3)--(c2);
\draw(b3)--(c3);

    \end{tikzpicture}
    \caption{Example of colouring in Theorem \ref{thm:rlffvs}}
    \label{fig:my_label}
\end{figure}
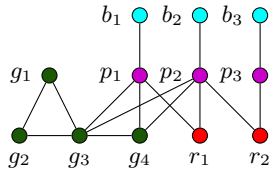

\begin{red}\label{Red3}\rm 
If there is a  purple vertex $p$ with more than one blue  neighbour
then reduce the number of blue neighbours  to one. 
\end{red}

\begin{figure}
\centering
\begin{tikzpicture}[scale=0.5]

\node[circle,fill=magenta, draw, inner sep=0 pt, minimum size=0.2cm](a0) at (0,0) [label=left:]{};

\node[circle,fill=cyan, draw, inner sep=0 pt, minimum size=0.2cm](a1) at (-2,1) [label=left:]{};
\node[circle,fill=cyan, draw, inner sep=0 pt, minimum size=0.2cm](a2) at (-1,1) [label=left:]{};
\node[circle,fill=cyan, draw, inner sep=0 pt, minimum size=0.2cm](a3) at (1,1) [label=left:]{};

\draw(a0)--(a1);
\draw(a0)--(a2);
\draw(a0)--(a3);
\draw[dotted, thick](-.5,1)--(.5,1);

\node[circle,fill=magenta, draw, inner sep=0 pt, minimum size=0.2cm](b0) at (4,0) [label=left:]{};

\node[circle,fill=cyan, draw, inner sep=0 pt, minimum size=0.2cm](b1) at (4,1) [label=left:]{};

\draw(b0)--(b1);

\draw[->,line width=1pt] (2,.5) -- (3,.5);
\end{tikzpicture}
    \caption{An illustration for Reduction Rule UEDS \ref{Red3}.}
    \label{fig-red3}
\end{figure}
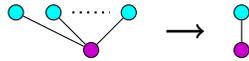

\noindent Notice that the  Reduction UEDS \ref{Red3} does not influence the set of feasible solutions  to the instance $(G,k)$. The fourth rule is based on the 
following observation. Suppose $G$ has a green vertex $g$ of degree $\geq 2k$. 
Suppose $N(g)=\{v_1,\ldots v_{2k}\}$. Note that each $v_i$ is either a green or purple vertex and hence of degree at least 2. We can easily construct  an edge dominating set 
$M$ with $|M|\geq k$ of $G$ without including any edge incident to $g$.

\begin{red}\label{Red4}\rm 
If  $G$ contains a green vertex $g$ of degree greater than or equal to 
$2k$, then conclude that we are dealing with a yes-instance.
\end{red}

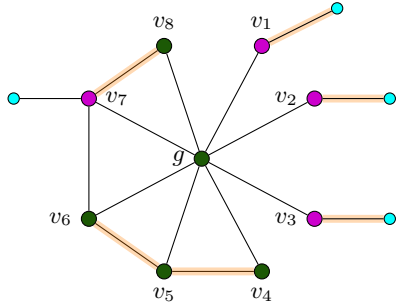
\begin{figure}
    \centering
\begin{tikzpicture}[scale=1]

\node[circle,fill=green, draw, inner sep=0 pt, minimum size=0.2cm](g) at (0,0) [label=left:$g$]{};

\node[circle,fill=magenta, draw, inner sep=0 pt, minimum size=0.2cm](v7) at (-1.5,.8) [label=right:$v_{7}$]{};
\node[circle,fill=cyan, draw, inner sep=0 pt, minimum size=0.15cm](b7) at (-2.5,.8) [label=left:]{};

\node[circle,fill=green, draw, inner sep=0 pt, minimum size=0.2cm](v6) at (-1.5,-.8) [label=left:$v_{6}$]{};
\node[circle,fill=green, draw, inner sep=0 pt, minimum size=0.2cm](v5) at (-.5,-1.5) [label=below:$v_{5}$]{};
\node[circle,fill=green, draw, inner sep=0 pt, minimum size=0.2cm](v4) at (.8,-1.5) [label=below:$v_{4}$]{};

\node[circle,fill=magenta, draw, inner sep=0 pt, minimum size=0.2cm](v2) at (1.5,.8) [label=left:$v_{2}$]{};
\node[circle,fill=cyan, draw, inner sep=0 pt, minimum size=0.15cm](b2) at (2.5,.8) [label=left:]{};

\node[circle,fill=magenta, draw, inner sep=0 pt, minimum size=0.2cm](v3) at (1.5,-.8) [label=left:$v_{3}$]{};
\node[circle,fill=cyan, draw, inner sep=0 pt, minimum size=0.15cm](b3) at (2.5,-.8) [label=left:]{};

\node[circle,fill=green, draw, inner sep=0 pt, minimum size=0.2cm](v8) at (-.5,1.5) [label=above:$v_{8}$]{};

\node[circle,fill=magenta, draw, inner sep=0 pt, minimum size=0.2cm](v1) at (.8,1.5) [label=above:$v_{1}$]{};
\node[circle,fill=cyan, draw, inner sep=0 pt, minimum size=0.15cm](b1) at (1.8,2) [label=left:]{};

\draw(g)--(v1);
\draw(g)--(v2);
\draw(g)--(v3);
\draw(g)--(v4);
\draw(g)--(v5);
\draw(g)--(v6);
\draw(g)--(v7);
\draw(g)--(v8);
\draw(v7)--(v8);
\draw(v7)--(v6);
\draw(v6)--(v5);
\draw(v5)--(v4);

\draw[draw, orange, opacity=0.3, line width=3](v7)--(v8);
\draw[draw, orange, opacity=0.3, line width=3](v5)--(v4);
\draw[draw, orange, opacity=0.3, line width=3](v6)--(v5);
\draw[draw, orange, opacity=0.3, line width=3](v1)--(b1);
\draw[draw, orange, opacity=0.3, line width=3](v2)--(b2);
\draw[draw, orange, opacity=0.3, line width=3](v3)--(b3);

\draw(v7)--(b7);
\draw(v1)--(b1);
\draw(v2)--(b2);
\draw(v3)--(b3);

\end{tikzpicture}
    \caption{Illustration of Reduction Rule UEDS \ref{Red4}. Note that $d(g)=8$ and $d(v_{i})\geq 2$ for all $1\leq i \leq 8$. A minimal edge dominating set of size $6$ that does not contain any edge incident to $g$, is shown here in orange colour.}
    \label{fig:red4}
\end{figure}

\noindent The fifth rule is based on the observation that having $k$ blue vertices implies that there is a matching of size at least $k$. We can construct a minimal edge dominating set of $G$ which contains all the edges of the matching. This implies that we have a yes-instance.

\begin{red}\label{Red5}\rm
If $G$ contains at least $k$ blue vertices,
then conclude that we are dealing with a yes-instance.
\end{red}

\noindent Next, we present a reduction rule that applies when $G$ has some red vertices. 

\begin{red}\label{Red6}\rm 
If $V_R$ is non-empty, then remove $V_R$ from $G$. 
The new instance is $(G- V_r,k)$.
\end{red}
 We make some important observations before we prove the correctness of this 
 reduction rule.  Let $p$ be a purple vertex and for simplicity let $N_B(p)=\{b\}, N_R(p)=\{r_1,r_2\}$
  and $N_G(p)=\{g_1,g_2\}$.
\begin{lemma}\label{obs1}\rm 
For every purple vertex $p$, at least one edge incident to $p$ is included in the solution. 
\end{lemma}
\proof Every purple vertex $p$ is adjacent to a blue vertex $b$. The only way to dominate the  edge $(p,b)$ is to  include to the solution either the edge itself  or  
another edge incident to $p$.\qed

\begin{lemma}\label{obs2}\rm 
 If a purple-red edge $(p,r)$
 is included in the solution, then
no other edges incident to $p$ can be included in the solution.
\end{lemma}

\begin{figure}[ht]
    \centering
    \begin{tikzpicture}[scale=0.8]
\node[circle,fill=green, draw, inner sep=0 pt, minimum size=0.2cm](c2) at (1,0) [label=below:$g_1$]{};
\node[circle,fill=green, draw, inner sep=0 pt, minimum size=0.2cm](c3) at (2,0) [label=below:$g_2$]{};
\node[circle,fill=red, draw, inner sep=0 pt, minimum size=0.2cm](c4) at (3,0) [label=below:$r_1$]{};
\node[circle,fill=red, draw, inner sep=0 pt, minimum size=0.2cm](c5) at (4,0) [label=below:$r_2$]{};
\node[circle,fill=magenta, draw, inner sep=0 pt, minimum size=0.2cm](b2) at (2,1) [label=left:$p'$]{};
\node[circle,fill=magenta, draw, inner sep=0 pt, minimum size=0.2cm](b3) at (3,1) [label=left:$p$]{};
\node[circle,fill=magenta, draw, inner sep=0 pt, minimum size=0.2cm](b4) at (4,1) [label=left:$p''$]{};

\node[circle,fill=cyan, draw, inner sep=0 pt, minimum size=0.2cm](a1) at (2,2) [label=left:$b'$]{};
\node[circle,fill=cyan, draw, inner sep=0 pt, minimum size=0.2cm](a2) at (3,2) [label=left:$b$]{};
\node[circle,fill=cyan, draw, inner sep=0 pt, minimum size=0.2cm](a3) at (4,2) [label=left:$b''$]{};

\draw(c2)--(b2);
\draw(a1)--(b2);
\draw(a2)--(b3);
\draw(a3)--(b4);
\draw(c3)--(b2);
\draw(c4)--(b3);
\draw(c5)--(b4);
\draw(c2)--(b3);
\draw(c4)--(b2);
\draw(c5)--(b3);
\draw(b3)--(c3);
\draw(c4)--(b4);
\draw(c2)--(c3);
    \end{tikzpicture}
    \caption{An illustration  for Lemma \ref{obs2}.}
    \label{fig-lemma3}
\end{figure}
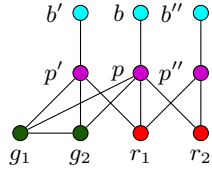

\proof Suppose $(p,r_{1})$ is included in the solution, 
then we show that $(p,r_{2})$ cannot be included in the solution. 
Assume, for the sake of contradiction,  that both  $(p,r_{1})$ and $(p,r_{2})$
 are included in the solution $M$. Notice that the edges incident to $p$ are
 not private edges, as they are dominated by two 
 edges $(p,r_{1})$ and $(p,r_{2})$ of $M$.  By Lemma \ref{private-edge-lemma}, every edge of $M$ has a private edge. Therefore the private edge of $(p,r_1)$  must be $(r_1,p')$ for some purple vertex $p'\neq p$. This is impossible, as by Lemma \ref{obs1}, the purple vertex $p'$ has an edge $e'$ incident to $p'$ in the solution. Therefore $(r_{1},p')$ is not a private edge and hence $(p,r_1)$ does not have
any private edge, a contradiction. \qed

\begin{lemma}\label{obs3}\rm  If  a purple-blue $(p,b)$ edge
 is included in the solution, then no other edges incident to $p$ can be included in the solution.
\end{lemma}

\noindent The proof of this lemma is essentially the same as the proof of Lemma \ref{obs2},

\begin{lemma}\label{obs4}\rm 
If a purple-green $(p,g)$ edge  is included in the solution, then no other edge of the form purple-red or purple-blue incident to $p$ can be included in the solution.
\end{lemma}

Based on the above observations, we will construct a solution $S'$ from $S$ such that no edge between a purple vertex and a red vertex is inside the set $S$ and $|S'|\geq |S|$. Let us assume that $S$ contains edges of the form $(v_{i},w_{i})$ where $v_{i}$ is coloured purple and $w_{i}$ is coloured red. Due to observations \ref{obs2},\ref{obs3} and \ref{obs4}, we know that there is a unique edge adjacent to $v_{i}$ inside the solution and that is $(v_{i},w_{i})$. We replace every edge $(v_{i},w_{i})$ in $S$ by an edge $(v_{i},z_{i})$ where $z_{i}$ is a unique blue neighbour of $v_{i}$.

\begin{lemma}
Reduction Rule \ref{Red6} is safe. 
\end{lemma}
\proof In one direction, we show that if $M$ is a solution to $(G,k)$ then
$M'$ is a solution to $(G-V_r,k)$, where $M'$ is obtained from $M$ be 
replacing every purple-red edge $(p,r) \in M$ by the purple-blue edge $(p,b)$. 
First we prove that $M'$ is an edge dominating set in $G-V_r$. 
Assume, for the sake of contradiction,
that $M'$ is not an edge dominating set in $G-V_r$, that is, an edge 
$e$ in $G-V_r$ is not dominated by $M'$ but it was dominated by an edge $e'$
in $M$. Clearly, $e'$ must be a purple-red edge $(p,r)$. Therefore, one endpoint 
of  $e$ must be $p$. Note that $r$ cannot be an endpoint of $e$ 
as it an edge in $G-V_r$. As $e'=(p,r)$ is replaced by $(p,b)$ in $M'$, and one endpoint 
of $e$ is $p$, $M'$ dominates $e$, a contradiction. Therefore, $M'$ is an edge-dominating 
set in $G-V_r$. Now, we claim that $M'$ is a minimal edge dominating set in $G-V_r$. 
Consider an edge $e$ in $M'\cap M$. \\

\noindent{\it Case 1.} Suppose $e=(p,g)$ where $p$ is a purple vertex and $g$ is a green vertex. By Lemma \ref{obs4},  if $e=(p,g)$ is in the solution, no
 purple-red $(p,r)$ or purple-blue edges $(p,b)$ can be included in the solution. Then, clearly 
 $(p,b)$ is a private edge of $e$ and it remains private edge of $e$ in $M'$ too. \\

\noindent{\it Case 2.} Suppose $e=(p,b)$ where $p$ is a purple vertex and $b$ is a blue vertex. By Lemma \ref{obs3},  if  the purple-blue edge $(p,b)$
 is included in the solution, then
no other edges incident to $p$ can be included in the solution.
Note that $e=(p,b)$  is its own private edge in both $M$ and $M'$.\\

\noindent{\it Case 3.}  Suppose $e=(g_1,g_2)\in M$ or $e=(p_1,p_2)\in M$.
It can be verified that $e$ will have a private edge in $M'$ as well.\\

\noindent Next consider an edge $e \in M'\setminus M$. By construction, $e$ must be of the 
form $(p,b)$ which has replaced  some purple-red edge $(p,r)$ in $M$. 
By Lemma \ref{obs3}, if  the purple-blue $(p,b)$ edge
 is included in the solution, then
no other edges incident to $p$ can be included in the solution.
Therefore,  no other edge in $M'$ dominates the edge $(p,b)$ but itself. 

For the other direction, let  $M$ be a minimal edge dominating set of $G-V_R$. 
We claim that $M$ is  a minimal edge dominating set in $G$. It is true because we get 
$G$ from $G-V_R$ by introducing some red vertices and some purple-red edges.
By Lemma \ref{obs1},
for every purple vertex $p$, at least one edge incident to $p$ is included in the solution. 
Therefore, $M$ dominates all edges of $G$ including the newly introduced 
purple-red edges. \qed

\noindent We claim the final reduction rule that explicitly bounds the size of the 
kernel.
\begin{red}\label{Red7}\rm 
Let $(G,k)$ be an input instance such that Reductions UEDS 1 to UEDS 6 are not applicable to $(G,k)$. If  $G$ has more than $4k^2-2$ vertices, then conclude that we are dealing with a yes instance. 
\end{red}

\noindent Since we cannot apply Reductions UEDS 1 to UEDS 6 anymore on graph $G$, we have 
$|V(G)|=|V_B|+|V_P|+|V_G|$. 
By Reduction Rule \ref{Red5},
if $|V_B|\geq k$,
then conclude that we are dealing with a yes-instance.
Therefore, we can assume that there are less than $k$ blue vertices. It also implies that there are less than $k$ purple vertices. Next, we note that 
if $G$ has a  minimal vertex cover of size  $2k$ then $(G,k)$ is a yes- instance. 
This is true because we can obtain a maximal matching of $G$ of size 
greater than or equal to $k$ by a greedy algorithm. 
Since the maximal matching is of size more than or equal to $k$ then we are done as it is also a minimal edge dominating set. The endpoints of the maximal matching forms a vertex cover of $G$. Now, consider the graph induced by the set of green vertices in $G$. Since we computed a minimal vertex cover of size less than $2k$, it implies that  $G[V_G]$ has a 
vertex cover $S$ of size less than $2k$.  Since we cannot apply Reduction UEDS \ref{Red1} anymore on $G$, $G[V_G]$ has no isolated vertices. Thus every vertex of $G[V_G]-S$ should be adjacent to some vertex from $S$. By Reduction UEDS \ref{Red4}, every vertex of $G[V_G]$ has degree less than $2k$. It follows that $|V_G-S|< 2k|S|$ and hence 
$|V_G|< (2k+1)|S|$. Since $|S|<2k$, we have $|V_G|< (2k+1)2k$. 
Therefore, we have $|V(G)|=|V_B|+|V_P|+|V_G|\leq (k-1)+(k-1)+(2k)(2k-1)=4k^2-2$, which concludes that  the Reduction Rule UEDS \ref{Red7} is safe.\\

Finally, we remark that all the reduction rules are trivially applicable in linear time. 
Thus we obtain the following theorem.

\begin{theorem}\label{thm:rlffvs}
    {\sc Upper Edge Dominating Set}  parameterized by the solution size $k$ admits a kernel of size $4k^{2}-2$.
\end{theorem}

\section{Single exponential time algorithm parameterized by solution size}

In this section, we construct a $2^{\mathcal{O}(k)}$ running time algorithm using dynamic programming parameterized by solution size. Assume we are given a 
{\sc Upper Edge Dominating Set} instance $(G,k)$. We find a maximal 
matching $M$ in $G$. Note that a maximal matching in $G$ is also a minimal edge dominating set 
in $G$. If
$|M|\geq k$, then we can clearly conclude that $(G,k)$ is a yes-instance, so assume
otherwise, that is, $|M|<k$. The  endpoints of the edges in $M$ constitute a 
vertex cover of $G$ of size at most $2k-2$.
As $G$ has a vertex cover of size bounded by $2k-2$, we can 
construct a tree decomposition of $G$ with width at most $2k-1$.
Therefore, if {\sc Upper Edge Dominating Set} can be solved in 
time $2^{\mathcal{O}(\omega)}$ where $\omega$ is the treewidth of the input graph,
then it can be solved in time $2^{\mathcal{O}(k)}$, where $k$ is the solution size. 
  We now prove the following theorem:
  
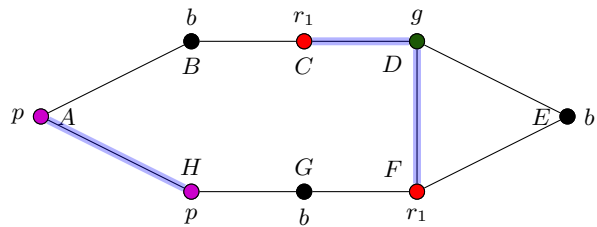
\begin{figure}
    \centering
\begin{tikzpicture}[scale=1]

\node[circle,fill=black, draw, inner sep=0 pt, minimum size=0.2cm](B) at (0,0) [label=above:$b$]{};
\node[circle,fill=black, draw, inner sep=0 pt, minimum size=0.2cm](B0) at (0,0) [label=below:$B$]{};

\node[circle,fill=red, draw, inner sep=0 pt, minimum size=0.2cm](C) at (1.5,0) [label=below:$C$]{};
\node[circle,fill=red, draw, inner sep=0 pt, minimum size=0.2cm](C0) at (1.5,0) [label=above:$r_{1}$]{};

\node[circle,fill=green, draw, inner sep=0 pt, minimum size=0.2cm](D) at (3,0) [label=below left:$D$]{};
\node[circle,fill=green, draw, inner sep=0 pt, minimum size=0.2cm](D0) at (3,0) [label=above:$g$]{};

\node[circle,fill=magenta, draw, inner sep=0 pt, minimum size=0.2cm](H) at (0,-2) [label=above:$H$]{};
\node[circle,fill=magenta, draw, inner sep=0 pt, minimum size=0.2cm](H0) at (0,-2) [label=below:$p$]{};

\node[circle,fill=black, draw, inner sep=0 pt, minimum size=0.2cm](G) at (1.5,-2) [label=below:$b$]{};
\node[circle,fill=black, draw, inner sep=0 pt, minimum size=0.2cm](G0) at (1.5,-2) [label=above:$G$]{};

\node[circle,fill=red, draw, inner sep=0 pt, minimum size=0.2cm](F) at (3,-2) [label=below:$r_{1}$]{};
\node[circle,fill=red, draw, inner sep=0 pt, minimum size=0.2cm](F0) at (3,-2) [label=above left:$F$]{};

\node[circle,fill=magenta, draw, inner sep=0 pt, minimum size=0.2cm](A) at (-2,-1) [label=right:$A$]{};
\node[circle,fill=magenta, draw, inner sep=0 pt, minimum size=0.2cm](A0) at (-2,-1) [label=left:$p$]{};

\node[circle,fill=black, draw, inner sep=0 pt, minimum size=0.2cm](E) at (5,-1) [label=right:$b$]{};
\node[circle,fill=black, draw, inner sep=0 pt, minimum size=0.2cm](E0) at (5,-1) [label=left:$E$]{};

\draw(A)--(B);
\draw(B)--(C);
\draw(C)--(D);
\draw(D)--(E);
\draw(E)--(F);
\draw(F)--(G);
\draw(G)--(H);
\draw(H)--(A);
\draw(D)--(F);

\draw[draw, blue , opacity=0.3, line width=3](C)--(D);
\draw[draw, blue , opacity=0.3, line width=3](F)--(D);
\draw[draw, blue , opacity=0.3, line width=3](A)--(H);

\end{tikzpicture}
    \caption{The blue edges shows a minimal edge dominating set of above graph. Note that the edges incident to black vertices are not contained in the solution. Edge $(A,H)$ forms a $K_{1,1}$ and its endpoints are purple. Edges $(C,D)$ and $(D,F)$ forms a $K_{1,2}$ where $D$ is colored green and $C,D$ are colored $r_{1}$.}
    \label{fig:tw}
\end{figure}  
  
  \begin{theorem}\label{treewidth}\rm 
  Given an $n$-vertex graph $G$ and its nice tree decomposition $T$ of width at most $\omega$, the 
  size of a maximum {\sc Upper Edge Dominating Set} of $G$ can be computed in time $45^{\omega}\cdot n^{\mathcal{O}(1)}$. 
  \end{theorem}
  
\proof 
Let $(T, \{X_{t}\}_{t \in V(T)})$ be a nice tree decomposition rooted at node $s$ 
of the input graph $G$.  For a node $t$ of $T$, let $V_t$ be the union of all
bags present in the subtree of $T$ rooted at $t$, including $X_t$. 
With each node $t$ of the tree decomposition we associate a subgraph 
$G_t=(V_t,E_t)$ where $$E_t=\{e~:~ e \mbox{ is introduced in the subtree rooted at } t\}.$$
A \emph{coloring} of bag $X_{t}$ is a mapping $f : X_{t} \rightarrow \{b,p,g,r_{0},r_{1}\}$ assigning five different colors to vertices of the bag. We give 
intuition  behind the  five colours. Suppose $A_t\subseteq E_t$ is a minimal 
edge dominating set of $G_t$. 
It is easy to observe that  subgraph $(V_t,A_t)$ is the disjoint union of 
isolated vertices  and  stars, isomorphic to $K_{1,r}$ for some $r\geq 1$.  Isolated vertices are coloured 
black; endpoints of $K_{1,1}$ are colored purple; 
 the internal node of a star graph $K_{1,r}$, with $r\geq 2$, is colored green and the $r$ leaves are colored red. See Fig. ?, which provides an illustration of the colouring.
 \begin{itemize}
    \item {\bf Black}, represented by $b$. The meaning is that the edges incident to 
    black vertices are not contained  in the partial solution in $G_t$.
    \item {\bf Purple}, represented by $p$. 
    If two adjacent vertices $u$ and $v$ are coloured purple then $e=(u,v)$ is contained
    in the partial  solution in $G_t$ but
    edges in $N[e]$ are not contained in the partial solution.  
    \item {\bf Green}, represented by $g$. For every green vertex $u$, at least two edges 
    incident to $u$ are contained  in the partial solution.
    \item {\bf Red}, represented by $r_0$ and $r_1$. 
    The meaning is that exactly one edge $(u,v)$ incident to every red vertex $u$ is 
    contained in the partial solution, where $v$ has to be a green vertex. 
    Furthermore,  vertices coloured $r_1$ must have at least one black neighbour where as
    vertices coloured $r_0$  have no black neighbours. 
\end{itemize}

\noindent For each node $t$ of $T$, we construct a table $dp_t(f,{\bf{y}}, n_r, n_{r_1},n_c, \alpha,\beta)\in \{ \mbox{true, false}\}$
where $f$ is a colouring of the bag $X_t$; ${\bf y}$ is a vector of length $n$; $n_r$, $n_{r_1}$, and  $n_c$
are integers between $0$ and $n$; $\alpha$ and $\beta$ are integers between $0$ and $m$.
 The vector ${\bf{y}}$ is of length $n$  and $i$th coordinate of vector 
 ${\bf{y}}$ is 
 \[ y(i) = 
            \begin{cases}
                0 &\quad\text{if $v_i$ is incident to no edges in the partial solution in $G_t$}\\
                1 &\quad \text{if $v_i$ is incident to exactly one edge in the partial solution in $G_t$ }\\
                2 & \quad \text{if $v_i$ is incident to at least two edges  in the partial solution in $G_t$.}\\
            \end{cases}   
    \]
 We use $n_r$ to denote the number of  red vertices in $G_{t}$; 
 $n_{r_1}$ to denote the number of red vertices with a  black neighbour in $G_{t}$; 
 $n_c$ to denote the number of vertices in $V_{t}\setminus X_{t}$ which satisfies the coloring condition. 
 Note that a black vertex satisfies the coloring condition if the number of edges 
 incident to it from the partial solution in  $G_t$ is zero. 
  A purple or red vertex satisfies the coloring condition if the number of edges 
 incident to it from the partial solution in  $G_t$ is exactly one.  
  A green  satisfies the coloring condition
  if the number of edges incident to it from the partial solution in  $G_t$ is  greater than or equal to $2$.  Finally $\alpha$  denotes the number of edges 
  in the partial solution in $G_t$ and  $\beta$ denotes the number of edges such that both the endpoints are colored black. We set $dp_t(f,  {\bf{y}}, n_r, n_{r_1},n_c, \alpha, \beta)= \mbox{true}$ if and only if there exists  
  a subset $A_t\subseteq E_t$ such that: 
\begin{enumerate}
    \item $ n_r= |\{ v\in V_t~:~ f(v)\in \{r_0, r_1\}\}| $
    \item $n_{r_1}=|\{ v\in V_t~:~ f(v)= r_1\}|$
    \item $y(i)=|\{e \in A_t~:~ \text{$e$ is incident to $v_{i}$}\}|$ for all $i$.
    \item $n_c$ is number of vertices in $V_{t}\setminus X_{t}$ satisfying the coloring conditions stated above.
    \item $\alpha = |A_{t}|$
    \item $\beta = |\{(u,v)\in E_t~:~ f(u)=f(v)= b\}|$.
\end{enumerate}
Note that the size of a minimal edge dominating set in $G$ is  $\alpha$ for which we have
$dp_s(\emptyset, {\bf y}, n_r, n_{r_1}, |V(G)|,\alpha,0)$= true, $n_r=n_{r_1}$, 
$n_c=|V(G)|$ and $\beta=0$. This is because we have $G=G_s$, $X_s=\emptyset$, which means that for $X_s$ we have only one colouring, the empty function; $n_r=n_{r_1}$ because we 
need every red vertex to have a black neighbour; $n_c=|V(G)|$ because we want every vertex to satisfy the colouring condition; and finally $\beta=0$ because we do not want to have 
any edge with both the endpoints coloured black in the final solution.\\

\noindent In the following, we  compute all entries $dp_t(f,  {\bf{y}}, n_r, n_{r_1},n_c, \alpha, \beta)$
in a bottom-up manner. There are $5^{\omega}\cdot 3^{\omega} \cdot n^3 \cdot m^2=15^{\omega}\cdot n^{O(1)}$ possible tuples $(f, {\bf{y}}, n_r, n_{r_1},n_c, \alpha, \beta)$
at  each node $t$ of $T$.
Thus to prove Theorem \ref{treewidth}, it suffices to show that each 
entry $dp_t(f,  {\bf{y}}, n_r, n_{r_1},n_c, \alpha, \beta)$ can be computed in time 
$3^{\omega}\cdot n^{O(1)})$ time assuming that the entries for the children of 
$t$ are already computed. \\

 \noindent  Now we introduce some notations. Let $X\subseteq V$ and consider a colouring $f~:~X\mapsto \{b, p, g, r_0,r_1\}$.
	 For $\gamma \in \{1,0,\hat1\}$ and $v\in V(G)$ a new colouring  $f_{v \mapsto \gamma}:X \cup \{v\}\mapsto \{b, p, g, r_0,r_1\}$ is defined as follows:
	$$f_{v \mapsto \gamma}(x)=\begin{cases}
	f(x)                     & \text{when $x\neq v$}\\
	\gamma					 & \text{when $x=v$}\\
	\end{cases}$$
	Let $f$ be a colouring of $X$, then the notation  $f|_{Y}$ is used to denote the restriction of $f$ to $Y$, where $Y\subseteq X$.
We now proceed to present the recursive formulas for the values of 
$dp_t(f,  {\bf{y}}, n_r, n_{r_1},n_c, \alpha, \beta)$.\\

\noindent{\bf Leaf node:} For a leaf node $t$ we have that $X_t=\emptyset$. Hence there is only
one empty colouring. Observe that $dp_t(f,  {\bf{y}}, n_r, n_{r_1},n_c, \alpha, \beta)=$ true if and only if $f=\emptyset$, $y(i)=0$ for all $i$, $n_r=n_{r_1}=n_c=0$,
$\alpha=0$, and $\beta=0$. These conditions can be checked in $O(1)$ time.\\

 \noindent{\bf Introduce vertex node:}  Suppose $t$ is an introduction node with a child $t^{\prime}$ such
 that $X_t=X_{t^{\prime}} \cup \{v_i\}$ for some $v_i\notin X_{t^{\prime}}$.
 Recall that we have not introduced any edges adjacent to $v_i$, so $v_i$ is isolated 
 in $G_t$.  For any colouring $f$ of $X_t$, we consider the following cases:\\
 
  \noindent{\it Case (i):} Let $f(v_i)\in \{b,g,p\}$. 
 Then, $dp_t(f,  {\bf{y}}, n_r, n_{r_1},n_c,\alpha, \beta)= \mbox{true}$ if and only if  $dp_{t'}(f|_{X_{t'}},  {\bf{y}}, n_r, n_{r_1},n_c, \alpha, \beta)= \mbox{true}$.\\

 \noindent{\it Case (ii):} Let $f(v_i)=r_{0}$. 
 Then, $dp_t(f,  {\bf{y}}, n_r, n_{r_1},n_c, \alpha,\beta)= \mbox{true}$
 if and only if  $dp_{t'}(f|_{X_{t'}},  {\bf{y}}, n_r-1, n_{r_1},n_c, \alpha,\beta )= \mbox{true}$.  \\
 
  \noindent{\it Case (iii):} Let $f(v_i)=r_1$. 
 Then, $dp_t(f,\cdot,\cdot,\cdot,\cdot,\cdot,\cdot)= \mbox{false}$. This is because
 we need to be sure that we do not introduce an isolated vertex with color $r_1$; 
 an isolated vertex cannot have a black neighbour. \\
 
 \noindent Therefore, $dp_t(f,{\bf{y}}, n_r, n_{r_1},n_c, \alpha,\beta)$ can be computed 
 in $O(1)$ time.\\
 
\noindent{\bf Introduce edge node:}  Suppose $t$ is an introduction edge node 
labeled with  edge $v_iv_j$ and let $t'$ be the child of $t$. 
Let $f$ be a coloring of $X_t$.  We consider two cases:\\
 
\noindent{\it Case (i):} Let $(v_i,v_{j}) \in A_{t}$. 
 Then, $dp_t(f,{\bf{y}}, n_r, n_{r_1},n_c, \alpha,\beta)= \mbox{true}$ if and only if 
 $dp_{t'}(f',  {\bf{y'}}, n_r, n'_{r_1},n_c, \alpha', \beta')= \mbox{true}$,
 where  \\
 
 \begin{enumerate}
 \item  $f(v_i)=r_1$, if $f'(v_i)=r_0$ and $f'(v_j)=b$;  $f(v_k)=f'(v_k)$ for 
 all $k\neq i$. 
\item  the $k$th coordinate of vector ${\bf y}$ is
      \[ y(k) = 
            \begin{cases}
              y'(k)+1  &\quad\text{for $k\in \{i,j\}$}\\
              y'(k) &\quad\text{otherwise}\\
            \end{cases}   
    \]
 \item       \[ n_{r_1} = 
            \begin{cases}
              n'_{r_1}+1  &\quad\text{if $f(v_{i})=r_0$ and $f(v_j)=b$}\\
              n'_{r_1} &\quad\text{otherwise}\\
            \end{cases}   
    \]

    \item $\alpha = \alpha'+1$
    \item     \[ \beta = 
            \begin{cases}
              \beta' +1  &\quad\text{if $f(v_{i})=b$ and $f(v_{j})=b$}\\
              \beta' &\quad\text{otherwise}\\
            \end{cases}   
    \]
 \end{enumerate}
 
\noindent{\it Case (ii):} Let $(v_i,v_{j}) \notin A_{t}$. 
 In this case $dp_t(f,  {\bf{y}}, n_r, n_{r_1},n_c, \alpha,\beta)= \mbox{true}$ if and only if 
 $dp_{t'}(f',  {\bf{y}}, n_r, n'_{r_1},n_c, \alpha, \beta')= \mbox{true}$,
 where $f$, $n_{r_1}$ and $\beta$ satisfy the above recurrence relation.  \\
 
 \noindent Therefore, $dp_t(f,{\bf{y}}, n_r, n_{r_1},n_c, \alpha,\beta)$ can be computed 
 in $O(k)$ time. \\

\noindent{\bf Forget node:}  Let $t$ be a forget node with a child $t'$ such
 that $X_t=X_{t'}\setminus \{v_i\}$ for some $v_i\in X_{t'}$. Then, 
 $dp_t(f,  {\bf{y}}, n_r, n_{r_1},n_c, \alpha,\beta)= \mbox{true}$ if and only if 
 $dp_{t'}(f_{v_i\rightarrow \gamma},  {\bf{y}}, n_r, n_{r_1},n'_c, \alpha, \beta)= \mbox{true}$ for 
 some $\gamma\in\{b,p,g,r_0,r_1\}$,
 where 
   \[ n_c = 
            \begin{cases}
              n'_c+1  &\quad\text{if $f(v_{i})=b,y(i)=0$}\\
              n'_c+1  &\quad\text{if $f(v_{i})=g,y(i)\geq 2$}\\
             n'_c+1  &\quad\text{if $f(v_{i})\in \{p, r_{0},r_{1}\},y(i)=1$}\\
             n'_c  &\quad\text{otherwise}
            \end{cases}   
    \]
 \noindent Therefore, $dp_t(f,{\bf{y}}, n_r, n_{r_1},n_c, \alpha,\beta)$ can be computed 
 in $O(1)$ time. \\ 
 
\noindent{\bf Join node:}  Let $t$ be a join node with children $t_{1}$ and $t_{2}$. 
Recall that $X_t=X_{t_1}=X_{t_2}$. Then,
$dp_t(f,{\bf{y}}, n_r, n_{r_1},n_c, \alpha,\beta)= \mbox{true}$
 if and only if there exist  $(f^1,{\bf{y^1}}, n^1_r, n^1_{r_1},n^1_c, \alpha^1,\beta^1)$ and 
 $(f^2,{\bf{y^2}}, n^2_r, n^2_{r_1},n^2_c, \alpha^2,\beta^2)$ such that\\
 $dp_{t_1}(f^1,{\bf{y^1}}, n^1_r, n^1_{r_1},n^1_c, \alpha^1,\beta^1)= dp_{t_2}(f^2,{\bf{y^2}}, n^2_r, n^2_{r_1},n^2_c, \alpha^2,\beta^2)= \mbox{true}$, where
 \begin{enumerate}
    \item $f(v_i)=f^1(v_i)=f^2(v_i)$ for all $v_i\in X_t$
    \item $y(i)=y^1(i)+y^2(i)$ for all $1\leq i \leq n$
    \item $n_r=n^1_r+n^2_r - |\{ u \in X_t~:~f(u)\in\{r_{0},r_{1}\}\}|$
    \item $n_{r_1}=n^1_{r_1}+ n^2_{r_1} - |\{ u\in X_t~:~f(u)=r_{1}\}|$
     \item $n_c=n^1_c+n^2_c$
    \item $\alpha=\alpha^1+\alpha^2$
     \item $\beta=\beta^{1}+\beta^{2}$
\end{enumerate}
There are at most $3^{\omega}$ possible pairs for $({\bf y^1,y^2})$ as ${\bf y^2}$ is uniquely 
determined by ${\bf y^1}$, $n$ possible pairs for $(n^1_r,n^2_r)$, for  $(n^1_{r_1},n^2_{r_1})$,
and for $(n^1_c,n^2_c)$; and $m$ possible pairs for $(\alpha^1,\alpha^2)$, and for 
$(\beta^1,\beta^2)$. In total there are $3^{\omega} n^{O(1)}$ candidates. Each candidate can be
checked in $O(1)$ time.  Therefore, $dp_t(f,{\bf{y}}, n_r, n_{r_1},n_c, \alpha,\beta)$ can be computed  in $3^{\omega} n^{O(1)}$ time.  \\

\noindent Since we assume that the number of nodes in a nice tree decomposition is $O(kn)$, the algorithm
requires $45^{\omega} \cdot n^{O(1)}$ time. At the root node $s$, we look at all records such that $dp_s(\emptyset, {\bf y}, n_r, n_{r_1}, n_c,\alpha,\beta)$= true,
 $n_r=n_{r_1}$, $n_c=|V(G)|$, $\beta=0$. The  maximum size of a minimal edge dominating set  is the maximum $\alpha$ satisfying $dp_s(\emptyset, {\bf y}, n_r, n_{r}, |V(G)|,\alpha,0)$= true.
 \qed

\section{Conclusion}

We proved that {\sc Upper EDS} admits a kernel of size $4k^{2}-2$ on general graphs. We have also provided a single exponential FPT algorithm when parameterized by treewidth which also provides a single exponential FPT algorithm parameterized by solution size. This algorithm is obtained by constructing a dynamic programming on graphs with bounded treewidth. We list some problems emerge from the results here:
(1) It remains open whether {\sc upper EDS} on general graphs admits a linear kernel or a matching lower bound can be proved,
(2) Can we improve the base of a single exponential FPT algorithm parameterized by treewidth?


\bibliographystyle{abbrv}
\bibliography{bibliography}
 
\end{document}